\documentclass[pre,superscriptaddress,showpacs,preprint]{revtex4-1}
\usepackage{amsmath,graphicx,amssymb,amsfonts}
\begin{document}

\title{ An Expression for the Granular Elastic Energy}
\author{Yimin Jiang}
\author{Hepeng Zheng}
\author{Zheng Peng}
\author{Liping Fu}
\affiliation{Central South University, Changsha, China 410083}
\author{Shixiong Song}
\author{Qicheng Sun}
\affiliation{State Key Laboratory for Hydroscience and Engineering, Tsinghua University,
Beijing, China 100084}
\author{Michael Mayer}
\author{Mario Liu}
\affiliation{Theoretische Physik, Universit\"{a}t T\"{u}bingen, 72076 T\"{u}bingen,
Germany}
\date{\today }

\begin{abstract}
Granular Solid Hydrodynamics (GSH) is a broad-ranged continual mechanical
description of granular media capable of accounting for static stress
distributions, yield phenomena, propagation and damping of elastic waves, the
critical state, shear band, and fast dense flow. An important input of GSH
is an expression for the elastic energy needed to deform the grains. The
original expression, though useful and simple, has some draw-backs.
Therefore, a slightly more complicated expression is proposed here that
eliminates three of them: (1)~The maximal angle at which an inclined layer
of grains remains stable is increased from $26^\circ$ to the more realistic
value of $30^\circ$. (2)~Depending on direction and polarization, transverse
elastic waves are known to propagate at slightly different velocities. The
old expression neglects these differences, the new one successfully
reproduces them. (3)~Most importantly, the old expression contains only the
Drucker-Prager yield surface. The new one contains in addition those named
after Coulomb, Lade-Duncan and Matsuoka-Nakai -- realizing each, and
interpolating between them, by shifting a single scalar parameter.
\end{abstract}

\maketitle

\section{Introduction}

GSH (brief for: \emph{Granular Solid Hydrodynamics}) is a continual
mechanical theory~\cite{granR2} constructed to account for a broad range of
granular phenomena, including static stress distribution~\cite%
{ge-1,ge-2,granR1}, incremental stress-strain relation~\cite{SoilMech},
yield~\cite{JL2}, propagation and damping of elastic waves~\cite{ge4},
elasto-plastic motion~\cite{JL3}, the critical state~\cite{critState}, shear
band and fast dense flow~\cite{denseFlow}. An important input of GSH is an
expression for the elastic energy needed to deform the grains~\cite{JL1}.
The energy density $w(u_{ij})$ is a function of the {elastic
strain} $u_{ij}$, which we define as the long-scaled, coarse-grained measure of how
and how much the grains are deformed. Therefore, both the elastic energy $w$
and the stress $\sigma_{ij}$ (which comes about only because the grains are
deformed) are necessarily functions of $u_{ij}$. Moreover, we have
\begin{equation}  \label{sigma-ij}
\sigma_{ij}=-\partial w/\partial u_{ij},
\end{equation}
because $\sigma_{ij}$ is closely related to the force, and $u_{ij}$ to the
coordinate. So  $\sigma_{ij}$ is given if $w$ is.

The relation between the elastic and total strain $\varepsilon_{ij}$ is only
simple at small increments, where $\delta
u_{ij}\approx\delta\varepsilon_{ij} $ holds. More generally, it is given by the evolution equation of $u_{ij}$~\cite{JL3,ge-1,ge-2}, and not by a function, because only $u_{ij}$ is a state variable, not $\varepsilon_{ij}$. As it
turns out, the critical state is simply the stationary solution of $u_{ij}$%
's evolution equation~\cite{critState}.

The expression $w(u_{ij})$ is an input of GSH. It may either be deduced
microscopically, or obtained iteratively in a trial and error process, by
comparing the ramification of the proposed expression with experimental
observations. As the first method is notoriously difficult, we choose trial
and error. The original expression,
\begin{equation}
w=\mathcal{B}\sqrt{\Delta }\left( {2}\Delta ^{2}/{5}+u_{s}^{2}/{\xi }\right),
\label{w0}
\end{equation}
is very simple, and a function of only two invariants, $\Delta \equiv
-u_{kk} $ and $u_{s}^{2}\equiv u_{ij}^{\ast }u_{ij}^{\ast }$. ($u_{ij}^{\ast
}\equiv u_{ij}-u_{kk}\delta _{ij}/3$ is the traceless or deviatoric strain.)
$\mathcal{B}$ and $\xi$ are two (density dependent) elastic coefficients.
The dependence on the third invariant $u_{t}^{3}\equiv
u_{ik}^{\ast}u_{kj}^{\ast }u_{ji}^{\ast }$ is neglected. Nevertheless, a
number of granular features are contained in this expression. First of all,
the measured velocity of elastic waves~\cite{jia} are well rendered~\cite%
{ge4}. Second, satisfactory agreement was achieved~\cite{SoilMech} with the
incremental stress-strain relation as observed and reported in~\cite{Kuwano}. In both cases, one is looking at small increment of the elastic strain, $%
\delta u_{ij}\approx\delta\varepsilon_{ij}$, and the calculation employing
Eq~(\ref{w0}) is purely elastic (or hyperelastic):
\begin{equation}
\delta\sigma_{ij}=\frac{\partial\sigma_{ij}}{\partial u_{k\ell}}\delta
u_{k\ell}=-\frac{\partial^2 w}{\partial u_{ij}\partial u_{k\ell}}%
\delta\varepsilon_{k\ell}.
\end{equation}
Third, the Drucker-Prager yield surface, requiring a granular system at rest
to have a ratio of shear stress $\sigma_s\equiv\sqrt{\sigma^*_{ij}%
\sigma^*_{ij}}$ and pressure $P\equiv\sigma_{\ell\ell}/3$ smaller than a
certain value, $\sigma_s/P<$~constant, is, as explained next, an integral part of this
expression.

Generally speaking, in a space spanned by stress components, there is a
surface that divides two regions in any granular media, one in which the
grains necessarily move, another in which they may be at rest. This surface
is usually referred to as the yield surface. Aiming to make its definition
more precise, we take the yield surface to be the divide between one region
in which elastic solutions are stable, and another in which they are not --
clearly, the medium may be at rest for a given stress only if an appropriate
elastic solution is stable. Since the elastic energy of a solution
satisfying the equilibrium condition $\nabla_j\sigma_j=0$ is an extremum~%
\cite{granR2}, the elastic energy is convex and minimal in the stable
region, concave and maximal in the unstable one. In the latter case, an
infinitesimal perturbation suffices to destroy the solution. The elastic
energy of Eq~(\ref{w0}) is convex only for
\begin{equation}  \label{2b-3}
u_s/\Delta\le\sqrt{2\mathcal{B}/\mathcal{A}}, \quad \text{implying}\quad
\sigma_s/P_\Delta\le\sqrt{2\mathcal{A}/\mathcal{B}}.
\end{equation}

In this paper, we propose a slight generalization of the energy, by
including the third invariant $u_{t}^{3}\equiv
u_{ik}^{\ast}u_{kj}^{\ast}u_{ji}^{\ast}$ and its elastic coefficient $\chi$,
as
\begin{equation}
w=B\sqrt{\Delta }\left( \frac{2}{5}\Delta ^{2}+\frac{1}{\xi }u_{s}^{2}-\frac{%
\chi }{\xi }\frac{u_{t}^{3}}{\Delta }\right).  \label{w}
\end{equation}
We shall in the rest of the paper examine its ramifications.

There are many different yield surfaces in the literature. The first was
proposed by Coulomb, who observed that slopes of sand piles never exceed a
critical value and saw an analogy with the friction law~\cite{Coulomb},
\begin{equation}
\frac{\sigma _{3}-\sigma _{1}}{\sigma _{3}+\sigma _{1}}=\sin \varphi ,
\label{C}
\end{equation}%
where $\varphi $ is the internal friction angle,  a material parameter, and $%
\sigma _{1}\leq \sigma _{2}\leq \sigma _{3}$ are eigenvalues of $\sigma
_{ij} $, ordered by their magnitude. Although the Coulomb yield model is
widely used for estimating the stability of granular materials at rest~\cite%
{Nedderman}, a number of other models are also frequently used by engineers, depending
on personal preference and experience, especially those by Drucker and Prager%
\cite{DP}, Lade and Duncan~\cite{LD}, Matsuoka and Nakai~\cite{MN}, given
respectively as
\begin{eqnarray}
\frac{\sigma _{s}}{P}&=&\frac{\sqrt{6}\sin \varphi }{\sqrt{3+\sin
^{2}\varphi } },  \label{DP} \\
\frac{\sigma _{1}\sigma _{2}\sigma _{3}}{27 P^{3}}&=&\frac{\left( 1-\sin
\varphi \right) \cos ^{2}\varphi }{\left( 3-\sin \varphi \right) ^{3}},
\label{LD} \\
\frac{\left( \sigma _{1}-\sigma _{3}\right) ^{2}}{\sigma _{1}\sigma _{3}}+
\frac{\left( \sigma _{2}-\sigma _{3}\right) ^{2}}{\sigma _{2}\sigma _{3}}&+&
\frac{\left( \sigma _{1}-\sigma _{2}\right) ^{2}}{\sigma _{1}\sigma _{2}}
=8\tan ^{2}\varphi.  \label{MN}
\end{eqnarray}%
As we shall see below, all three yield models are also
reproduced in satisfactory approximation by the new energy expression. In
addition, it also improves on other residual discrepancies between the
elastic theory and measurements, such as in the incremental stress-strain
data, or the speed of elastic waves. All this indicates that the new energy,
now with three parameters: $B,\xi ,\chi $, is capable of giving a more
accurate description of granular elasticity.

We note that the new cubic term in Eq~(\ref{w}) was first introduced by
Humrickhouse in~\cite{Humrickhouse}, in an attempt to increase the maximum
angle $\theta _{\max }$ of inclination, at which a granular layer remains
stable. Unfortunately, he only considered negative values of $\chi$. As
these did not yield any improvement, he abandoned this term. As we shall see
in section II, a positive $\chi$ does yield a larger $\theta _{\max }$, of
around $30^\circ$. In section III, we deduce the yield surface associated
with Eq~(\ref{w}); in section IV, incremental stress-strain relation and
granular acoustics are studied. All support a positive $\chi $. Section V
contains discussion and conclusions.

\section{The Maximum angle of inclination}

We consider noncohensive granular materials, with constant mass density $%
\rho $. Denoting
\begin{eqnarray}
A_{1} &\equiv &u_{xx}\text{, }A_{2}\equiv u_{yy}\text{, }A_{3}\equiv u_{zz}%
\text{, }  \notag \\
A_{4} &\equiv &u_{xy}\text{, }A_{5}\equiv u_{xz}\text{, }A_{6}\equiv u_{yz},
\label{10}
\end{eqnarray}
the $6\times 6$ Hessian matrix of the function $w(A_{\alpha })$ is
\begin{equation}
H_{\alpha \beta }=-\frac{\partial ^{2}w}{\partial A_{\alpha }\partial
A_{\beta }},  \label{9}
\end{equation}
with $\alpha ,\beta =1,2,...,6$ and the eigenvalues $h_{1}\leq h_{2}\leq
\cdots \leq h_{6}$. The associated yield surface, written as a function of
the stress components, is given by
\begin{equation}
h_{1}=0.  \label{h1=0}
\end{equation}
As the eigenvalues frequently lack analytic expressions, we consider instead
its determinant, $\det H_{\alpha \beta }=0$. This is of course only a
necessary condition, as we need to ensure that the vanishing eigenvalue is
the smallest one, $h_1$, while the other ones are larger. This is done
numerically.

\begin{figure}[tbh]
\begin{center}
\includegraphics[scale=0.9]{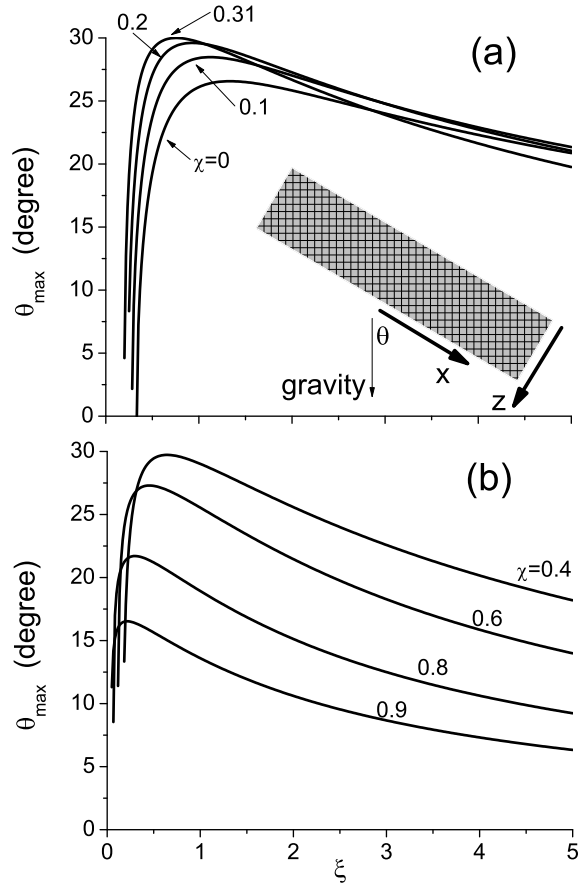}
\end{center}
\caption{Variation of the maximum angle of inclination with the parameters $%
\protect\xi $ and $\protect\chi $ calculated with the potential Eq~(\protect\ref%
{w}) and the strain Eq~(\protect\ref{INC-uij}).}
\label{fig1}
\end{figure}

Next, consider an infinite granular layer in gravity, inclined by an angle $%
\theta $, see Fig.\ref{fig1}. The elastic strain is taken to assume the form~%
\cite{JL1,Humrickhouse}
\begin{equation}
u_{ij}=\Delta \left(
\begin{array}{ccc}
0 & 0 & \tau  \\
0 & 0 & 0 \\
\tau  & 0 & -1%
\end{array}%
\right) ,  \label{INC-uij}
\end{equation}%
with $\tau \equiv u_{xz}/\Delta $, where $\tau ^{2}=u_{s}^{2}/\left( 2\Delta
^{2}\right) -1/3$. Inserting the strain~Eq~(\ref{INC-uij}) into (\ref{sxz},\ref%
{szz-1}), we obtain
\begin{eqnarray}
\sigma _{xz} &=&-{\frac{2+\chi }{\xi }\Delta }^{3/2}\tau \,,  \label{INC-sxz}
\\
\sigma _{zz} &=&\frac{9\xi +5\chi +15}{9\xi }{\Delta }^{3/2}+\frac{\chi +2}{%
2\xi }{\Delta }^{3/2}{\tau }^{2}.  \label{INC-szz}
\end{eqnarray}%
In addition, because of the force balance $\nabla _{j}\sigma _{ij}=\rho g_{i}
$, the two stresses are also related by
\begin{equation}
\sigma _{xz}=\sigma _{zz}\tan \theta .  \label{INC-forceBalance}
\end{equation}%
It is worth remarking that if we insert the strain Eq~(\ref{INC-uij}) into the
Hessian matrix $H_{\alpha \beta }$, we find for $\chi >0$ that two
eigenvalues are $4-\chi \left( 1\pm 3\,\sqrt{{1}+4\,{\tau }^{2}}\right) $,
implying that a stable layer exists only for $\chi <1$, because at least one
eigenvalue is negative for $\chi >1$. As the other eigenvalues cannot be
expressed analytically, we now consider $\det (H_{\alpha \beta })=0$,
finding
\begin{equation}
\det (H_{\alpha \beta })=\frac{4\left( 2+\chi \right) }{{\xi }^{6}}{\Delta }%
^{3}\,D_{1}D_{2},  \label{INC-det}
\end{equation}%
with%
\begin{eqnarray}
D_{1} &=&{\left( -4+2\,\chi +2\,{\chi }^{2}+9\,{\chi }^{2}{\tau }^{2}\right)
,}  \label{INC-D1} \\
D_{2} &=&{27\chi ^{2}\left( \chi +2\right) \tau ^{4}+3\left( -30\chi +12\chi
^{2}+5\chi ^{3}+36\xi \chi ^{2}+12\right) \tau ^{2}}  \label{INC-D2} \\
&&{+2\left( \chi -1\right) \left( 4\chi +\chi ^{2}-6\right) +36\,\xi
\,\left( {\chi }^{2}-1\right) .}  \notag
\end{eqnarray}%
The real roots of $D_{1}=0$ are ${\tau }_{\pm }{=\pm }\sqrt{4-2\chi
^{2}-2\chi }{/}3{\chi }$, that of $D_{2}=0$ are%
\begin{eqnarray}
\tau _{c} &=&\,\pm \frac{1}{3{\chi }}\sqrt{{\frac{k_{1}+\sqrt{k_{2}}}{%
2\left( 2+\chi \right) }}},\qquad \text{ \ with}  \label{INC-1} \\
k_{1} &=&-5\,{\chi }^{3}-12\left( 3\xi +1\right) \chi ^{2}+30\,\chi -12,
\label{INC-2} \\
k_{2} &=&\chi ^{6}-72\xi \chi ^{5}+12\left( 108\xi ^{2}-5\right) \chi
^{4}-24\left( 72\xi +11\right) \allowbreak \chi ^{3}  \label{INC-3} \\
&&+36\left( 48\xi +25\right) \chi ^{2}-720\chi +144.  \notag
\end{eqnarray}%
(Because $k_{1}<0$ for $\chi \in \left( 0,1\right) $, positive square root
for $\sqrt{k_{2}}$ ie taken here.) Requiring the vanishing eigenvalue to be
the smallest one, we find yield to occur at $\tau =\tau _{c}$, implying a
maximum angle of inclination by employing Eqs~(\ref{INC-sxz},\ref{INC-szz}),
\begin{equation}
\theta _{\max }=\mp \arctan \,\left[ {\frac{18\,\left( 2+\chi \right) \tau
_{c}}{9\left( \chi +2\right) \tau _{c}^{2}+18\,\xi +30+10\,\chi }}\right] .
\label{INC-4}
\end{equation}%
Inserting $\tau _{c}$ of Eq~(\ref{INC-1}) into it, and remembering that the
sign only indicates a left or right inclination, we obtain the absolute
value for the maximal angle as
\begin{equation}
\theta _{\max }=\arctan \,\left[ {\frac{6\,\sqrt{2}{\chi }\sqrt{2+\chi }%
\sqrt{{k_{1}+\sqrt{k_{2}}}}}{k_{1}+\sqrt{k_{2}}+36\,\xi {\chi }^{2}+60{\chi }%
^{2}+20\,\chi ^{3}}}\right].   \label{INC-5}
\end{equation}%
It is nice to have an analytic expression for $\theta _{\max }$, which
reduces to the results of~\cite{JL1} for $\chi \rightarrow 0$ i.e.: $\tau
_{c}\rightarrow \sqrt{\xi -1/3}$ and $\theta _{c}\rightarrow \arctan \left[
\sqrt{9\xi -3}/\left( 3\xi +2\right) \right] $.

In fig.\ref{fig1}, the expression of Eq~(\ref{INC-5}) for $\theta _{\max }$
is plotted against $\xi $ for various $0<\chi <1$. The biggest value $\theta
_{\max }=30^{\circ}$ is achieved at $\xi \simeq 0.71$ and $\chi \simeq 0.31$%
. More specifically, as $\chi $ increases from $0$ to $0.31$, the peak of $%
\theta _{\max }$ versus $\xi $ increases too, but decreases after that.

A final remark: the elastic strain of Eq~(\ref{INC-uij}) is a result of
assuming that displacement vector $U_{i}$ varies only with the layer depth $%
z $, and has nonvanishing components along the $x$ and $z$ directions, $%
U_{y}=0 $. Then yield, at $\theta =\theta _{\max }$,  occurs simultaneously
in the whole layer. This is an idealization. In reality, the displacement
vector should also have a nonvanishing $y$ component. Then yield will probably
start at the layer top. This case will be studied elsewhere.

\section{The Yield Surface}

In this section, we consider uniform stress $\sigma _{ij}$ and density $\rho$%
, and employ the coordinate system of the principle directions of the
stress. The elastic strain $u_{ij}$ is also uniform and in its principle
system, ie, $\sigma _{ij}=0$ and $u_{ij}=0$ for $i\neq j$, see Appendix.

Instead the principle strains $u_{xx}$, $u_{yy}$, $u_{zz}$, we shall, for
simplicity, use $\Delta $, $\mathcal{S}\geq 0$ and $L$, where
\begin{eqnarray}
u_{xx} &=&-\frac{\Delta }{3}\left[ 1-\mathcal{S}\sin \left( L-\frac{\pi }{3}%
\right) \right] ,  \label{uxx} \\
u_{yy} &=&-\frac{\Delta }{3}\left( 1+\mathcal{S}\sin L\right) ,  \label{uyy}
\\
u_{zz} &=&-\frac{\Delta }{3}\left[ 1-\mathcal{S}\sin \left( L+\frac{\pi }{3}%
\right) \right] .  \label{uzz}
\end{eqnarray}%
(In soil mechanics $L$ is usually called as Lode angle, of which value range
is taken as $[0,2\pi )$). And we have the following expressions for the
principle stresses,
\begin{eqnarray}
\frac{\sigma _{xx}}{{\Delta }^{3/2}} &=&3\,\sqrt{3}+{\frac{\mathcal{S}}{4\xi
}\left( 12\cos L-4\sqrt{3}\sin L+\sqrt{3}\mathcal{S}\right) }  \label{sxx} \\
&&{+}\frac{\chi \mathcal{S}^{2}}{24\xi }\left( 6\sqrt{3}\cos 2L-18\sin 2L+%
\sqrt{3}\mathcal{S}\sin 3L\right) ,  \notag \\
\frac{\sigma _{yy}}{{\Delta }^{3/2}} &=&3\sqrt{3}\,+\frac{\sqrt{3}}{4\xi }%
\mathcal{S}\left( \mathcal{S}+8\sin L\right)  \label{syy} \\
&&-\frac{\sqrt{3}\chi }{24\xi }\mathcal{S}^{2}\left( 12\cos 2L-\mathcal{S}%
\sin 3L\right) ,  \notag \\
\frac{\sigma _{zz}}{{\Delta }^{3/2}} &=&3\sqrt{3}\,-\frac{\mathcal{S}}{4\xi }%
\left( 12\cos L+4\sqrt{3}\sin L-\sqrt{3}\mathcal{S}\right)  \label{szz} \\
&&+\frac{\chi \mathcal{S}^{2}}{24\xi }\left( 18\sin 2L+6\sqrt{3}\cos 2L+%
\sqrt{3}\mathcal{S}\sin 3L\right) .  \notag
\end{eqnarray}%
The pressure and the deviatoric stresses $q,Q$ are simpler, with $\Delta $
factored out,
\begin{eqnarray}
P &\equiv &\sigma _{kk}/3=\frac{\sqrt{3}}{\xi }{\Delta }^{3/2}\left( 3\xi \,+%
\frac{1}{4}\mathcal{S}^{2}+\frac{\chi }{24}\mathcal{S}^{3}\sin 3L\right) ,
\label{P} \\
q &\equiv &\sigma _{zz}-\sigma _{xx}=\frac{3}{2\xi }{\Delta }^{3/2}\,%
\mathcal{S}\left( \chi \mathcal{S}\sin 2L-4\cos L\right) ,  \label{q} \\
Q/3 &\equiv &\left( P-\sigma _{yy}\right) =\frac{\sqrt{3}}{2\xi }\Delta
^{3/2}\mathcal{S}\left( \chi \mathcal{S}\cos 2L-2\sin L\right) .  \label{Q}
\end{eqnarray}

\subsection{The Cylindrically Symmetric Case}

For $\sigma _{xx}=\sigma _{yy}$ and $u_{xx}=u_{yy}$, the sample is
cylindrically symmetric, implying a Lode angle $L=\pi /6$ or $7\pi /6$, see
Eqs.(\ref{uxx},\ref{uyy}). Cylindrical symmetry is usually assumed in
analyzing the so-called "triaxial test," widely used in soil mechanics.
Inserting these Lode angles into Eqs~(\ref{P},\ref{q}), we obtain
\begin{equation}
\frac{q}{P}=\frac{18\,\mathcal{S}}{\chi \mathcal{S}^{3}+6\mathcal{S}%
^{2}+72\xi }\times \left\{
\begin{array}{c}
\left( \chi \mathcal{S}-4\right) \\
\left( \chi \mathcal{S}+4\right)%
\end{array}%
\right. \text{ \ }\left.
\begin{array}{l}
\text{for }L=\pi /6, \\
\text{for }L=7\pi /6.%
\end{array}%
\right.  \label{Tri-1}
\end{equation}%
Note $L=\pi /6$ is the case of triaxial extension with $q<0$, while $L=7\pi
/6$ is the case of triaxial compression with $q>0$. At yield, $\mathcal{S=S}%
_{yield}$, Eq~(\ref{Tri-1}) implies $q\sim P$, and in the stress space
spanned by $P,q$ yields two straight lines.

To obtain the value for $\mathcal{S}_{yield}$, we calculate the determinant $%
\det (H_{\alpha \beta })$, obtaining
\begin{equation}
\det (H_{\alpha \beta })=\frac{27{\Delta }^{3}}{16{\xi }^{6}}%
\,D_{1}^{2}D_{2}^{2}D_{3},  \label{Tri-det}
\end{equation}%
where for $L=\pi /6$:
\begin{eqnarray}
D_{1} &=&{\chi }\mathcal{S}{+2},  \label{Tri-2} \\
D_{2} &=&{\chi }\mathcal{S}{-4},  \label{Tri-3} \\
D_{3} &=&\chi ^{2}\mathcal{S}^{4}-8\chi \mathcal{S}^{3}-24\mathcal{S}%
^{2}-144\xi \chi \mathcal{S}+288\xi ;  \label{Tri-4}
\end{eqnarray}%
and for $L=7\pi /6$:
\begin{eqnarray}
D_{1} &=&-{\chi }\mathcal{S}{+2},  \label{Tri-5} \\
D_{2} &=&-{\chi }\mathcal{S}{-4},  \label{Tri-6} \\
D_{3} &=&\chi ^{2}\mathcal{S}^{4}+8\chi \mathcal{S}^{3}-24\mathcal{S}%
^{2}+144\xi \chi \mathcal{S}+288\xi .  \label{Tri-7}
\end{eqnarray}%
If $\chi =0$, the equation $\det (H_{\alpha \beta })=0$ reduces to $D_{3}=48%
\mathcal{S}^{2}-576\xi =0$ for both Lode angles, implying $\mathcal{S}%
_{yield}=\sqrt{12\xi }$ for extension and compression. Inserting this into
Eq~(\ref{Tri-1}), we have $q/P=\pm \sqrt{3/\xi }$, the two straight yield
lines are then symmetric about the P-axis.

\begin{figure}[bth]
\begin{center}
\includegraphics[scale=0.3]{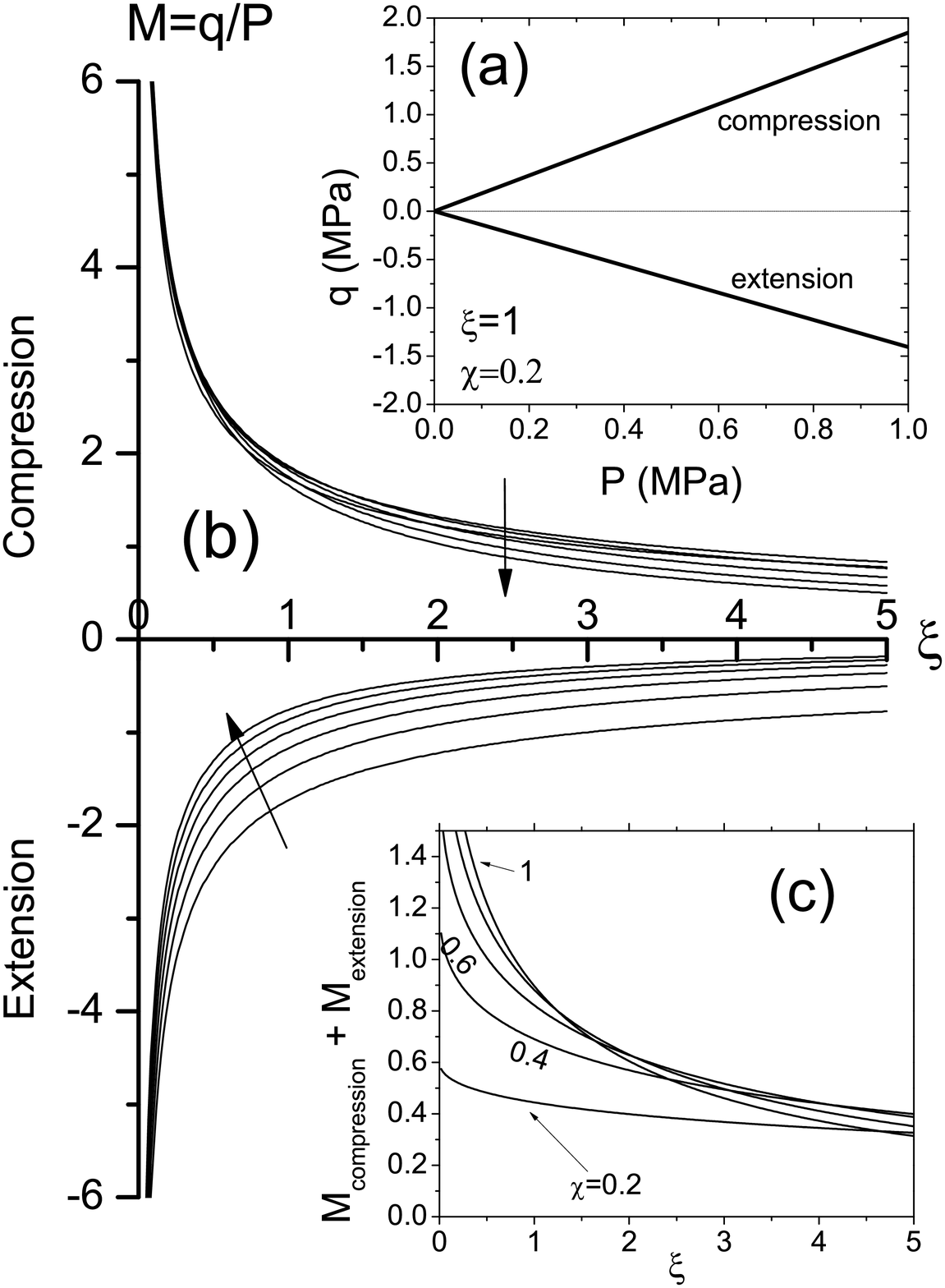}
\end{center}
\caption{Yield behavior as determined by the energy Eq~(\protect\ref{w}),
for the case of cylindrical symmetry. (a) Straight yield line for $\protect%
\xi=1$ and $\protect\chi=0.2$. (b) Variation of the slope of yield line with
$\protect\xi$ for $\protect\chi=0$,0.2,0.4,0.6,0.8,1 (increasing indicated
by arrows). (c) The difference between the slopes for compression and
extension.}
\label{fig2}
\end{figure}

For $\chi \neq 0$, one needs to resort to numerical calculation, which shows
that $\mathcal{S}_{yield}\geqslant 0$ is given by the smallest positive root
of $D_{3}=0$. Yet because Eqs~(\ref{Tri-4}) and (\ref{Tri-7}) are different,
$\mathcal{S}_{yield}$ is different for extension and compression, and so are
the two slopes, see fig.\ref{fig2}-a. This unsymmetric behavior of yield is
well documented and familiar in soil mechanics. Within the present
framework, it is a measure of how much the third invariant influences
granular elasticity. Fig.\ref{fig2}-b shows variation of the slope of yield
line with the parameter $\xi $, at varying $0\leqslant \chi <1$. In
agreement with the observations, the absolute value of the slopes is always
greater for compression than that for extension, see fig.\ref{fig2}-c.

\subsection{The General Case}

\begin{figure}[tbh]
\begin{center}
\includegraphics[scale=0.9]{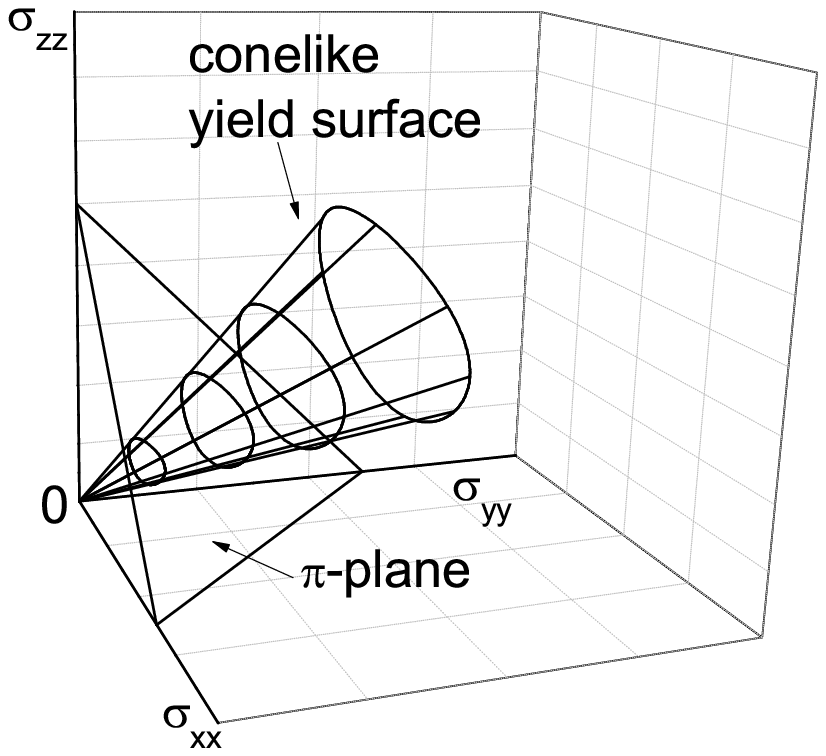}
\end{center}
\caption{Yield surface, $\protect\pi $-plane and their intersection loci in
principle-stress space. The $\protect\pi $-plane is defined by constant $%
\protect\sigma_{xx}+\protect\sigma_{yy}+\protect\sigma_{zz}$.}
\label{pi-plane}
\end{figure}
\begin{figure}[tbh]
\begin{center}
\includegraphics[scale=0.5]{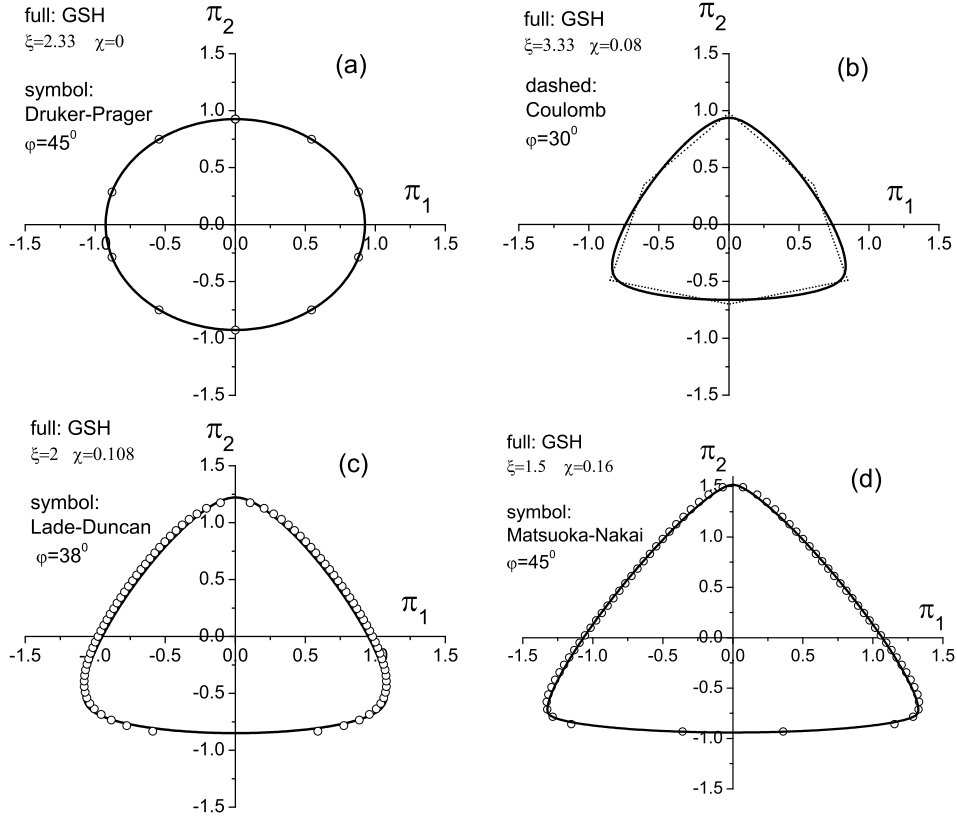}
\end{center}
\caption{Yield loci in the $\protect\pi $-plane calculated with the energy
Eq~(\protect\ref{w}), in comparison with various yield models, from (a) to
(d): Drucker-Prager, Coulomb, Lade-Duncan, Matsuoka-Nakai.}
\label{fig3}
\end{figure}

Next we consider the general case, with $\sigma _{xx}\neq \sigma _{yy}\neq
\sigma _{zz}$. Inserting the expressions (\ref{uxx},\ref{uyy},\ref{uzz})
into the Hessian matrix, and calculating its determinant, we have
\begin{equation}
\det (H_{\alpha \beta })=\frac{27\Delta ^{3}}{16\xi ^{6}}D_{1}D_{2},
\label{Tri-9}
\end{equation}%
where%
\begin{eqnarray}
D_{1} &=&\chi ^{3}\mathcal{S}^{3}\sin 3L-6\,{\chi }^{2}{\mathcal{S}}^{2}+32,
\label{Tri-10} \\
D_{2} &=&\chi ^{3}\mathcal{S}^{5}\sin 3L-6\,{\chi }^{2}{\mathcal{S}}%
^{4}-40\chi \mathcal{S}^{3}\sin 3L  \label{Tri-11} \\
&&-48\left( 3\xi \chi ^{2}+1\right) \mathcal{S}^{2}+576\,\xi .  \notag
\end{eqnarray}%
It can be demonstrated numerically that for any $L\in \left[ 0,2\pi \right) $%
, yield $\mathcal{S}=\mathcal{S}_{yield}$ is given by the smallest positive
root of $D_{1,2}=0$, which is also computed numerically. Inserting the yield
strain $\left( L,\mathcal{S}_{yield}\right) $ into (\ref{sxx},\ref{syy},\ref%
{szz}) the yield surface in the stress space can be plotted. Since the
compressional strain $\Delta $ is an overall factor, the surface is
conelike, as illustrated in fig.\ref{fig3}. We present the yield surface (as
is customary in soil mechanics) with a closed curve given by its
intersection with the so called $\pi $-plane, defined by $P=$const., see fig.%
\ref{pi-plane}. For the conelike surface it is convenient to introduce the
rectangular coordinates in the $\pi $-plane defined by
\begin{eqnarray}
\pi _{1} &=&\left( \sigma _{zz}-\sigma _{yy}\right) /\sqrt{2}P,
\label{Tri-12} \\
\pi _{2} &=&\left( 2\sigma _{xx}-\sigma _{yy}-\sigma _{zz}\right) /\sqrt{6}P.
\label{Tri-13}
\end{eqnarray}
Using Eqs~(\ref{sxx},\ref{syy},\ref{szz}), they become%
\begin{eqnarray}
\pi _{1} &=&\frac{6\sqrt{6}\mathcal{S}\left( \mathcal{S}\chi \sin 2L-4\cos
L\right) }{\chi \left( \sin 3L\right) \mathcal{S}^{3}+6\mathcal{S}^{2}+72\xi
},  \label{pi1} \\
\pi _{2} &=&\sqrt{\frac{3}{2}}\frac{6\mathcal{S}^{2}\chi \cos 2L-24\mathcal{S%
}\sin L+24\sqrt{3}\mathcal{S}\cos L-6\sqrt{3}\allowbreak \mathcal{S}^{2}\chi
\sin 2L}{\chi \left( \sin 3L\right) \mathcal{S}^{3}+6\mathcal{S}^{2}+72\xi }.
\label{pi2}
\end{eqnarray}%
$\allowbreak $Inserting $\left( L,\mathcal{S}_{yield}\right) $ into them the
yield curve in the $\pi $-plane can be plotted, see fig.\ref{fig3}. On the
other hand, the yield surfaces of the Coulomb, Drucker-Prager, Lade-Duncan
and Matsuoka-Nakai models, also conelike, and their loci in the $\pi $%
-plane, may be plotted directly -- using Eqs~(\ref{C},\ref{DP},\ref{LD},\ref%
{MN}) and Eqs~(\ref{Tri-12},\ref{Tri-13}) -- and compared.

The Drucker-Prager yield model can be reproduced exactly by the energy Eq~(%
\ref{w}) with $\chi =0$. It is a circle in the $\pi $-plane, see fig.\ref%
{fig3}-a. As $\chi $ increases from 0, the yield curve of Eq~(\ref{w}) is
distorted from a circle to a triangle-like curve, in fairly good agreement
with the Lade-Duncan and Matsuoka-Nakai models, see fig.\ref{fig3}-b,c.
Although the Coulomb model is a hexagon, with three more corners, see the
dotted line of fig.\ref{fig3}-b, the difference seems insignificant.

We note the large difference between the 
maximum angle of inclination $\theta _{\max }$ and  the friction
angle $\varphi $. Employing the parameters of Fig.6 below: $\xi =1$, $\chi =0.25$, we obtain $\theta _{\max }=29.8^\circ$, and $\varphi =65^\circ, 54^\circ, 60^\circ$ for the Coulomb, Lade-Duncan, and Matsuoka-Nakai model, respectively. 

\section{The Compliance Tensor}

Since GSH is a unified description of granular behavior, the energy of Eq~(%
\ref{w}) is expected to also account for the elastic stiffness and the speed
of elastic waves, both experimentally accessible. The stiffness tensor $%
M_{ijmn}$ is of fourth order and defined as
\begin{equation}
\delta\sigma _{ij}\equiv M_{ijmn}\delta u_{mn}=-(\partial ^{2}w/\partial
u_{ij}\partial u_{mn})\delta u_{mn},  \label{sound-1}
\end{equation}%
while the compliance tensor is its inverse,
\begin{equation}
du_{ij}=\lambda _{ijmn}d\sigma _{mn}.  \label{sound-2}
\end{equation}%
The differential forms of Eqs~(\ref{sound-1},\ref{sound-2}) are also
referred to as \emph{incremental stress-strain relation} in soil mechanics.
Physically, the components of the stiffness tensor can be interpreted as
response coefficients of a small stress change to a strain increment, or
vice versa for those of the compliance tensor.

\begin{figure}[t]
\begin{center}
\includegraphics[scale=0.3]{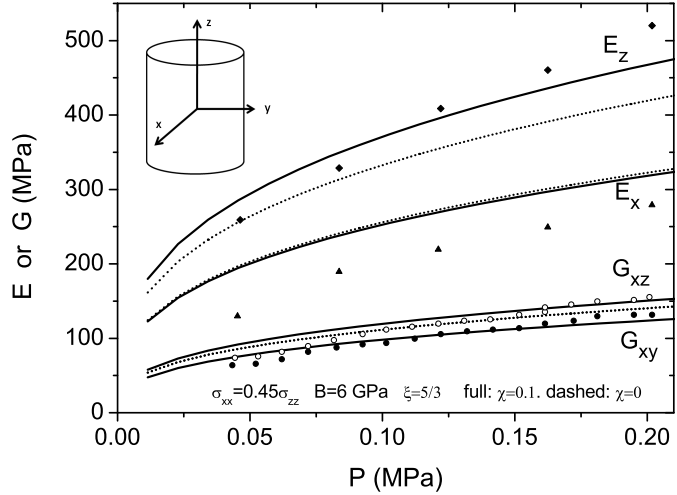}
\end{center}
\caption{Comparison of compliance coefficients between those measured in
\protect\cite{Kuwano} and calculated via Eq~(\protect\ref{sound-1}).}
\label{fig4}
\end{figure}

Systematic measurements of the response coefficients as functions of stress
and density was carried out by Kuwano and Jardine for the case of
cylindrical symmetry using a triaxial apparatus~\cite{Kuwano}. A comparison
of their data with the GSH calculation obtained for $\chi =0$ was given in~%
\cite{JL2,SoilMech}. Here we consider whether the agreement may be further
improved by a finite $\chi $. In fig.\ref{fig4}, we plotted the four
compliance components, denoted as $E_{xx}\equiv -1/\lambda _{xxxx}$, $%
E_{zz}\equiv -1/\lambda _{zzzz}$, $G_{xz}\equiv -1/4\lambda _{xzxz}$ and $%
G_{xy}\equiv -1/4\lambda _{xyxy}$, as functions of $P$. The agreement is
clearly improved for $\chi =0.1$. Especially noteworthy is the fact that $%
G_{xz}$ and $G_{xy}$ degenerate for $\chi =0$, but are split appropriately,
in the correct order, for $\chi=0.1$, again indicating that the sign of $%
\chi $ is positive.

\section{Elastic Waves}

\begin{figure}[bth]
\begin{center}
\includegraphics[scale=0.3]{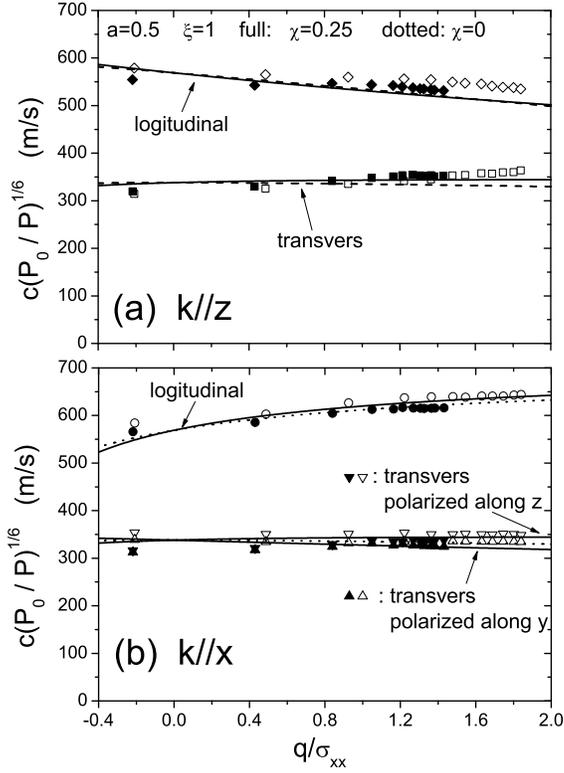}
\end{center}
\caption{Comparison of sound speeds as calculated employing the energy expression of Eq~(\protect\ref{w})
with the measurements reported in \protect\cite{jia}. $P_0$ is the
atmosphere pressure.}
\label{fig5}
\end{figure}

A further important elastic property is the speed $c_{i}$ of elastic waves,
which can be calculated from the square root of the appropriate eigenvalues
of the acoustic tensor, see~\cite{ge4, LL6},%
\begin{equation}
C_{ij}=\frac{k_{m}k_{n}}{\rho k^{2}}\frac{\partial ^{2}w}{\partial
u_{im}\partial u_{nj}}  \label{sound-matrix}
\end{equation}%
where $k_{m}$ is the wave vector of propagation, with $k^{2}\equiv
k_{m}k_{m} $. Inserting the energy of Eq~(\ref{w}) into (\ref{sound-matrix}%
), and eliminating the strain with the help of Eq~(\ref{sigma-ij}), we can
obtain different velocities as functions of the stress: $c_{i}=c_{i}\left(
\sigma _{ij}\right) $. Such a calculation with $\chi =0$ has been given in~%
\cite{ge4}, resulting in satisfactory agreement, though the weak variation
of the transverse wave velocity with varying polarization and direction of
propagation was not reproduced. Here, we focus on this and the additional
difference taking $\chi \neq 0$. Since the energy is a homogenous function
of the strain, or equivalently the stress, the stress dependence of the
velocities can be factorized as $c_{i}\sim P^{1/6}f\left( k_{m},\xi ,\chi
,\sigma _{ij}^{\ast}/P \right) $, where the factor $f\sim c_{i}/P^{1/6}$
represents the influences of shear stresses on sound speeds (from which the
values of the parameters $\xi ,\chi $ may be obtained with great accuracy).
Fig.~\ref{fig5} shows the calculated $c_{i}/P^{1/6}$, setting $\xi =1$ and $%
\chi =0.25$, in comparison with the data reported in~\cite{jia}. Cylindrical
symmetry is assumed, and the wave is taken to propagate along axial (Fig.~%
\ref{fig5}-a) and radial directions (Fig.~\ref{fig5}-b). Then $c_{i}/P^{1/6}$
depends only on $q/P$, or equivalently on $q/\sigma _{xx}=3(q/P)(3-q/P)^{-1}$%
. Clearly, GSH reproduces the data fairly well, where especially the order
of splitting of the transverse waves velocity, when either polarized along
the axial or the radial direction is correct, once again indicating that $%
\chi>0$.

Finally, we note that irrespective of the energy expression, there can only be two different transverse velocities in a cylindrical geometry along the principle axes. Therefore, if hardened experimental evidences to the contrary arises, this would indeed be a reason to include  an additional variable such as the intrinsic anisotropy.

Assuming a general expression for the energy,
$d\tilde{w}=\dots d\Delta+\dots du_{s}+\dots du_{t}$,
with
$d\Delta =-\delta_{ij}du_{ij}$, $du_{s}  =u_{ij}^{0}du_{ij}/{u_{s}}$,
$du_{t}  =u_{im}u_{mj}/{u_t ^{2}}$, we calculate the stress as
\begin{equation}
{\sigma}_{ij}=\tilde{f}_{1}(\Delta,u_{s},u_t )\delta_{ij}+\tilde{f}_{2}(\Delta,u_{s},u_t )u_{ij}+\tilde{f}_{3}(\Delta,u_{s},u_t )u_{im}u_{mj},
\end{equation}
where $\tilde{f}_{i}$ depend on the energy $\tilde w$.
Similarly, the stiffness tensor $\tilde{M}_{ijkl}$ is generally given as
\begin{multline}
\tilde{M}_{ijkl}=g_{1}\delta_{ij}\delta_{kl}+g_{2}\left(u_{ij}\delta_{kl}+u_{kl}\delta_{ij}\right)+g_{3}\left(u_{im}u_{mj}\delta_{kl}+u_{km}u_{ml}\delta_{ij}\right)\\
+g_{4}\left(\delta_{ik}\delta_{jl}+\delta_{il}\delta_{jk}\right)+g_{5}u_{ij}u_{kl}+g_{6}\left(u_{ij}u_{km}u_{ml}+u_{kl}u_{im}u_{mj}\right)\\
+g_{7}u_{im}u_{mj}u_{km}u_{ml}+g_{8}\left(u_{ki}\delta_{lj}+u_{kj}\delta_{li}+u_{li}\delta_{jk}+u_{lj}\delta_{ik}\right)\label{eq:SteifigkeitstensorAllgemein}
\end{multline}
with $g_{i}=g_{i}(\Delta,u_{s},u_t )$ again to be calculated from $\tilde w$.
In the system of principle axes,
$u_{ij}=u_{(i)}\delta_{ij}$, the stiffness tensor reads
\begin{eqnarray}
\tilde{M}_{ijkl} & = & \Bigl[g_{1}+g_{2}\left(u_{(i)}+u_{(k)}\right)+g_{5}u_{(i)}u_{(k)}+g_{3}\left(u_{(i)}^{2}+u_{(k)}^{2}\right)\nonumber \\
 &  & +g_{6}\left(u_{(k)}^{2}u_{(i)}+u_{(i)}^{2}u_{(k)}\right)\delta_{ij}+g_{7}u_{(i)}^{2}u_{(k)}^{2}\Bigr]\delta_{ij}\delta_{kl}\nonumber \\
 &  & \Bigl[g_{4}+g_{8}\left(u_{(i)}+u_{(j)}\right)\Bigr]\left(\delta_{ik}\delta_{jl}+\delta_{il}\delta_{jk}\right)\nonumber \\
 & := & g(i,k)\delta_{ij}\delta_{kl}+h(i,j)\left(\delta_{ik}\delta_{jl}+\delta_{il}\delta_{jk}\right).\label{eq:SteifigkeitstensorAllgemeinHaupt}
\end{eqnarray}
The acoustic tensor
$\tilde{C}_{il}\sim\tilde{M}_{ijkl}n_{j}n_{k}$ (with $n_i\equiv k_i/|k|$) therefore is
\begin{eqnarray*}
\tilde{C}_{il} & \sim & \left(g(i,l)+h(i,l)\right)n_{i}n_{l}\\
 &  & +\left(g_{4}+g_{8}\left(u_{(i)}+\left[u_{2}+\left(u_{1}-u_{2}\right)n_{1}^{2}+\left(u_{3}-u_{2}\right)n_{3}^{2}\right]\right)\right)\delta_{il}\\
 & = & H(i,l)n_{i}n_{l}+\left(g_{4}+g_{8}\left(u_{(i)}+u_{dev}\right)\right)\delta_{il},
\end{eqnarray*}
with $H(i,l)=g(i,l)+h(i,l)$ and $u_{dev}(\vec{n})=u_{2}+\left(u_{1}-u_{2}\right)n_{1}^{2}+\left(u_{3}-u_{2}\right)n_{3}^{2}$.

For cylindrical symmetry, denoting z as the preferred direction,  $u_{1}=u_{2}\neq u_{3}$,
we have $H(1,1)=H(2,2)=H(1,2)$ and $u_{dev}=u_{1}+\left(u_{3}-u_{1}\right)n_{3}^{2}$. We have, if the wave vector is along z ($n_{3}=1$, $n_{1}=n_{2}=0$),
\begin{eqnarray}
\tilde{C}_{z} & = & \left(\begin{array}{ccc}
g_{4}+g_{8}\left(u_{1}+u_{3}\right) & 0 & 0\\
0 & g_{4}+g_{8}\left(u_{1}+u_{3}\right) & 0\\
0 & 0 & H_{33}+g_{4}+2g_{8}u_{3}
\end{array}\right);\quad\label{eq:Kz}
\end{eqnarray}
if it is along x ($n_{1}=1$, $n_{2}=n_{3}=0$),
\begin{eqnarray}
\tilde{C}_{x} & = & \left(\begin{array}{ccc}
H_{11}+g_{4}+2g_{8}u_{1} & 0 & 0\\
0 & g_{4}+2g_{8}u_{1} & 0\\
0 & 0 & g_{4}+g_{8}\left(u_{3}+u_{1}\right)
\end{array}\right);\label{eq:Kx}
\end{eqnarray}
if it is along y ($n_{2}=1$, $n_{1}=n_{3}=0$),
\begin{eqnarray}
\tilde{C}_{y} & = & \left(\begin{array}{ccc}
g_{4}+2g_{8}u_{1} & 0 & 0\\
0 & H_{11}+g_{4}+2g_{8}u_{1} & 0\\
0 & 0 & g_{4}+g_{8}\left(u_{3}+u_{1}\right)
\end{array}\right).\label{eq:Ky}
\end{eqnarray}
We note two different eigenvalues of the acoustic tensor for the transverse case,
\begin{equation}
\tilde{C}_{z,11}=\tilde{C}_{z,22}=\tilde{C}_{x,33}=\tilde{C}_{y,33}\neq\tilde{C}_{x,22}=\tilde{C}_{y,11},
\end{equation}
and realize that there can only be two different velocities. Note also that without $g_{8}$ that stems from the third invariant, only one transverse velocitiy exists.

\section{Discussion and Conclusions}

Macroscopically speaking, granular media at rest are elastic, and one can
account for all its static, mechanic behavior including yield, compliance
coefficients, and sound propagation by a single potential, the elastic
energy. Finding an quantitatively appropriate potential is certainly
interesting and useful, and the expression given in Eq~ (\ref{w}) seems a
good starting point. It is a generalization of the potential given in~\cite%
{granR1,JL1}, with a term depending on the cubic strain invariant. All
empirical yield models and the other experiments considered above support
this term. As mentioned in the introduction, yield models in soil mechanics
are usually selected by personal experience, preference or convenience, and
there seems no consensus which one is best. The uncertainty of the situation
may be reduced with the help of the considered potential, as it unifies
these models, and link them to other easily measurable elastic properties
such as sound speeds, and compliance tensor. In this context, we would like
to stress the importance of more accurate and systematic experiments, such
as simultaneous measurement of sound velocity and yield.

It is important to realize that yield as considered here happens at the
highest possible stress at which the system may maintain an elastic, static
solution. It is different from that obtained under stead shear: When
approaching the critical state from an isotropic one, the approach is not
monotonous if the starting density is high, and the limiting, so-called
critical stress is lower than some of the stress values the system has
undergone. At all these stress states, if the strain rate is stopped, the
system will stop as well and retain its stress statically. Therefore, it is
still below yield at all these stress states. Measurements carried out with
steady shearing samples therefore do not reveal the yield surface.
Interpreting the results as such will lead to discrepancies.

Moreover, density also influences elastic properties of granular materials,
which was considered in~\cite{granR2}, but is neglected here for simplicity.
In~\cite{granR2}, we have shown how to include the density dependence of the
energy $w$ to account for the so-called ``cap"~\cite{cs}, or the sound
velocity as measured by Hardin and Richart~\cite{HR}. These features are
easily transferred to the energy of Eq~(\ref{w}). Summarizing, we believe
that although the details of granular elasticity is complicated, and more
accurate experiments are needed for further clarification, Eq~(\ref{w})
provides a reasonable start point.

\appendix
\section{Stress-Strain Relation\label{ssr}}

From the energy of Eq~(\ref{w}), we may calculate the stress as a function
of the elastic strain via $\sigma _{ij}=-\partial w/\partial u_{ij}$,
obtaining
\begin{align}
\sigma _{xy}& ={\frac{3\chi \left( u_{xz}u_{yz}-u_{xy}u_{zz}\right) -\left(
\chi +2\right) \Delta u_{xy}}{\sqrt{\Delta }\xi },}  \label{sxy} \\
\sigma _{xz}& ={\frac{3\chi \left( u_{xy}u_{yz}-u_{xz}u_{yy}\right) -\left(
\chi +2\right) \Delta u_{xz}}{\sqrt{\Delta }\xi },}  \label{sxz} \\
\sigma _{yz}& ={\frac{3\chi \left( u_{xy}u_{xz}-u_{yz}u_{xx}\right) -\left(
\chi +2\right) \Delta u_{yz}}{\sqrt{\Delta }\xi },}  \label{syz}
\end{align}%
and%
\begin{align}
q& \equiv \sigma _{zz}-\sigma _{xx}={\frac{1-\chi }{\xi /2}\sqrt{\Delta }%
\left( u_{xx}-u_{zz}\right) +\frac{3\chi }{\xi }}\frac{%
u_{yz}^{2}-u_{xy}^{2}-u_{xx}^{2}+u_{zz}^{2}}{\sqrt{\Delta }},  \label{s-q} \\
\widetilde{q}& \equiv \sigma _{yy}-\sigma _{xx}={\allowbreak \frac{1-\chi }{%
\xi /2}\sqrt{\Delta }\left( u_{xx}-u_{yy}\right) +}\frac{\allowbreak 3\chi }{%
\xi }\frac{u_{yz}^{2}-u_{xz}^{2}-u_{xx}^{2}+u_{yy}^{2}}{\sqrt{\Delta }}.
\label{s-Q}
\end{align}%
\begin{align}
\sigma _{zz}& =\frac{9\xi +5\chi +15\allowbreak }{9\xi }{\Delta }^{3/2}+%
\frac{3\left( \chi +2\right) \allowbreak }{2\xi }\sqrt{\Delta }\left(
u_{xx}+u_{yy}\right)  \label{szz-1} \\
& +\frac{1}{\sqrt{\Delta }\xi }\left(
u_{xx}^{2}+u_{xx}u_{yy}+u_{yy}^{2}+u_{xy}^{2}+u_{xz}^{2}+u_{yz}^{2}\right)
\notag \\
& +\frac{\chi }{2\xi \sqrt{\Delta }}\left(
u_{xx}^{2}+4u_{xx}u_{yy}+u_{yy}^{2}-2u_{xy}^{2}+u_{xz}^{2}+u_{yz}^{2}\right)
\notag \\
& +\frac{3\chi }{2{\Delta }^{3/2}\xi }\left[ \left(
u_{xy}^{2}-u_{xx}u_{yy}\right) \left( u_{yy}+u_{xx}\right)
-u_{xz}^{2}u_{yy}-u_{yz}^{2}u_{xx}+2u_{xy}u_{xz}u_{yz}\right]  \notag
\end{align}%
Clearly, if the strain is diagonal, the stress is too. In the principle
coordinate, the stress-strain relations becomes
\begin{align}
q& ={\frac{1-\chi }{\xi /2}\sqrt{\Delta }\left( u_{xx}-u_{zz}\right) +\frac{%
3\chi }{\xi }}\frac{-u_{xx}^{2}+u_{zz}^{2}}{\sqrt{\Delta }}, \\
\widetilde{q}& ={\allowbreak \frac{1-\chi }{\xi /2}\sqrt{\Delta }\left(
u_{xx}-u_{yy}\right) +}\frac{\allowbreak 3\chi }{\xi }\frac{%
-u_{xx}^{2}+u_{yy}^{2}}{\sqrt{\Delta }}.
\end{align}%
and%
\begin{align}
\sigma _{zz}& =\frac{9\xi +5\chi +15\allowbreak }{9\xi }{\Delta }^{3/2}+%
\frac{3\left( \chi +2\right) \allowbreak }{2\xi }\sqrt{\Delta }\left(
u_{xx}+u_{yy}\right) \\
& +\frac{1}{\sqrt{\Delta }\xi }\left(
u_{xx}^{2}+u_{xx}u_{yy}+u_{yy}^{2}\right)  \notag \\
& +\frac{\chi }{2\xi \sqrt{\Delta }}\left(
u_{xx}^{2}+4u_{xx}u_{yy}+u_{yy}^{2}\right)  \notag \\
& +\frac{3\chi }{2{\Delta }^{3/2}\xi }\left[ \left( -u_{xx}u_{yy}\right)
\left( u_{yy}+u_{xx}\right) \right]  \notag
\end{align}

\acknowledgments{This work is partly supported by the National Natural Science Foundation of
China (Grant No. 10904175).}


\begin{thebibliography}{99}

\bibitem{granR2} Y.M. Jiang and M. Liu, Granular Matter, \textbf{11}, 139
(2009); Y.M. Jiang and M. Liu, \emph{Mechanics of Natural Solids}, edited by
{D.}~{Kolymbas}{and}{G.}~{Viggiani}, {Springer}, pp. {27--46} (2009); {G.}~{%
Gudehus}, {Y.M.}~{Jiang}, {and} {M.}~{Liu}, {Granular Matter}, \textbf{1304}%
, {319} (2011).

\bibitem{ge-1} D.O. Krimer, M. Pfitzner, K. Br\"{a}uer, Y. Jiang, M. Liu,
\emph{Granular Elasticity: General Considerations and the Stress Dip in Sand
Piles,} Phys. Rev. \textbf{E74}, 061310 (2006).

\bibitem{ge-2} K. Br\"{a}uer, M. Pfitzner, D.O. Krimer, M. Mayer, Y. Jiang,
M. Liu, \emph{Granular Elasticity: Stress Distributions in Silos and under
Point Loads,} Phys. Rev. \textbf{E74}, 061311 (2006);

\bibitem{granR1} Y.M. Jiang, M. Liu, \emph{A Brief Review of ``Granular
Elasticity",} Eur. Phys. J. \textbf{E~22,} 255 (2007).

\bibitem{SoilMech} Y.M. Jiang, M. Liu, \emph{Incremental stress-strain
relation from granular elasticity: Comparison to experiments,} Phys. Rev.
\textbf{E 77}, 021306 (2008).

\bibitem{JL2} Y.M. Jiang, M. Liu, \textit{Energy Instability Unjams Sand and
Suspension,} Phys. Rev. Lett. \textbf{93}, 148001(2004).

\bibitem{ge4} {M.}~{Mayer} and {M.}~{Liu}, {Phys. Rev. E} {\bf 82}, {042301 } ({2010}).

\bibitem{JL3} Y.M. Jiang, M. Liu, \emph{From Elasticity to Hypoplasticity:
Dynamics of Granular Solids,} Phys. Rev. Lett. \textbf{99}, 105501 (2007).

\bibitem{critState}
{S.}~{Mahle}, {Y.}~{Jiang}, and {M.}~{Liu},
{arXiv:1006.5131v3[physics.geo-ph]} ({2010}).

\bibitem{denseFlow}
{Stefan~Mahle},
{Yimin~Jiang}, {M.~Liu},
{arXiv:1010.5350v1  [cond-mat.soft]} ({2010}).

\bibitem{JL1} Y.M. Jiang, M. Liu, \emph{Granular Elasticity without the
Coulomb Condition,} Phys. Rev. Lett. \textbf{91}, 144301 (2003).

\bibitem{jia} Y. Khidas and X.P. Jia, Phys. Rev. E, \textbf{81}, 021303
(2010).

\bibitem{Kuwano} R. Kuwano and R. J. Jardine, G\'{e}otechnique 52, 727 2002.

\bibitem{Humrickhouse} P.W. Humrickhouse, J.P. Sharpe,1 and M.L. Corradini,
\emph{Comparison of hyperelastic models for granular materials}, Phys.Rev.
\textbf{E 81}, 011303 (2010)

\bibitem{Coulomb} C.A. Coulomb, Mem. de Math. de l'Acad. Royale des Science
7, (1776) 343.

\bibitem{Nedderman} R.M. Nedderman, Statics and Kinematics of Granular
Materials, Cambridge university press,  Cambridge (1992).

\bibitem{DP} D.C. Drucker and W Prager, Soil mechanics and plastic analysis
for limit design. Quarterly of Applied Mathematics, 10(2), 157 (1952).

\bibitem{LD} P.V. Lade and J.M. Duncan, Cubic triaxial tests on
cohensionless soil, Proc. ASCE, JSMFD, vol 99, No SM10, 1973; P.V. Lade and
J.M. Duncan, Elastoplastic Stress-Strain Theory for Cohensionless Soil,
Proc. ASCE, JGTD, vol 101, No GT10 (1975).

\bibitem{MN} H. Matsuoka, On the significance of the spatial mobilized
plane. Soils \& Foundations. 1976 6(1): 91-100; H. Matsuoka and T. Nakai,
Stress-Strain relationship of soil based on the SMP. Proc. 9th ICSMFE,
specialty session 9, 153-163 (1977).

\bibitem{LL6} L.D. Landau and E.M. Lifshitz, \textit{Theory of Elasticity}
(New York, Pergamon Press, 3rd edn. 1986)

\bibitem{cs} A. Schofield and P. Wroth, Critical State Soil Mechanics.
McGraw-Hill, London (1968)

\bibitem{HR} B.O. Hardin, F.E. Richart, Elastic wave velocities in granular
soils. J. Soil Mech. Found. Div. ASCE 89(SM1), 33--65 (1963)
\end{thebibliography}
\end{document}